# An approximate analytical (structural) superposition in terms of two, or more, α-circuits of the same topology: Pt. 2 – the "internal circuit mechanism"

*Emanuel Gluskin*

The Galilean Academic College, Holon Institute of Technology, Holon 58102, *and* Electrical Engineering Department of the Ben-Gurion University, Beer-Sheva 84105, Israel.
gluskin@ee.bgu.ac.il.

**Abstract**: This is the second part, after [1], of the research devoted to analysis of 1-ports composed of similar conductors ("*f*-circuits") described by the characteristic $i = f(v)$ of a polynomial type. This analysis is performed by means of the power-law "α-circuits" introduced in [2], for which $f(v) \sim v^\alpha$. The *f*-circuits are constructed from the α-circuits of the same topology, with the proper $\alpha$, so that the given topology is kept, and '*f*' is an additive function of the connection. Explaining the situation described in detail in [1], we note and analyze a simple "circuit mechanism" that causes *the difference between the input current of the f-circuit and the sum of the input currents of the α-circuits before the composition to be relatively small*. The case of two degrees, $f(v) = D_m v^m + D_n v^n$, $m \neq n$ is treated in the main proofs. Some simulations are presented, and some boundaries for the error of the superposition are found. The cases of $f(.)$ being a polynomial of the third or fourth degrees are finally briefly considered.

## 1. Introduction

The present work treats, after [1] the "*f*-circuit" shown in Fig. 1, which is a 1-port of arbitrary structure composed of similar conductors.

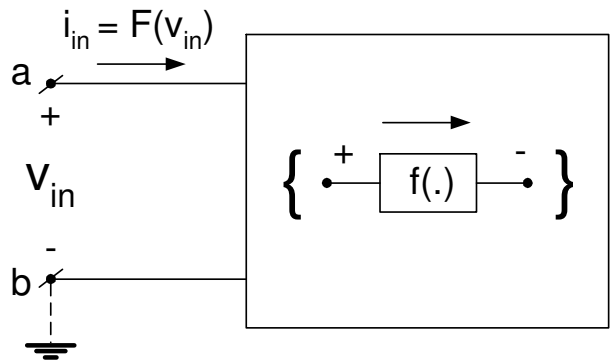

Fig.1: The 1-port (the "*f*-circuit") of a given topology, composed of similar conductors $f(.)$. Magnetic and dielectric d.c. realizations are also possible. The most typical case below is when '*f*' is a two-term polynomial, and we use the term "polynomial circuit". For the one-term (power-law) $f(v) \sim v^\alpha$, $\alpha > 0$; we speak about an $f_\alpha$-circuit, or "α-circuit". The α-circuits are the "building blocks" in our constructions.

We widely use (this is eq. (3) in [1]) the special case of

$$f(.) = f_\alpha(.) = D \cdot (.)^\alpha \qquad (1)$$

recalling [1,2] that $D$ does not influence the nodal potentials $v_k$ and (thus) the voltage drops $v_s$ on the elements, i.e. $D$ is just the factor for the actual currents, including the input current $F(.)$.

For $f$-circuits of the same topology and having the same input source, work [1] introduced a specific "$f$-connection", such that '$f$' is an additive variable of the connection. Namely, short-circuiting of all of the respective nodes (Fig. 2) yields a 1-port of the same topology, with the '$f(.)$' summed. It is necessary, however, to see the initially given circuits (now, generally, not 1-ports) in the connection.

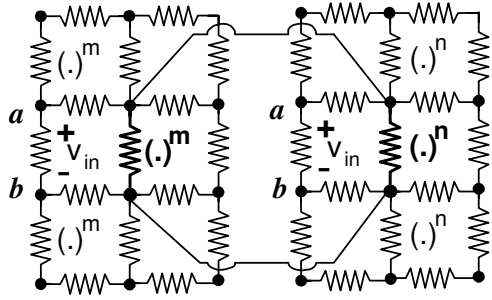

Fig. 2: The node-to-node "$f$-connection" of two circuits of the same topology, here one with $D_m(.)^m$ and another with $D_n(.)^n$. *All* the respective nodes are connected in pairs, as is shown for four nodes. Obviously, the resulting circuit is of the same topology, and $f(.) = f_1(.) + f_2(.)$, here $D_m(.)^m + D_n(.)^n$. Before the connection, the composing circuits are named "$f_m$-circuit" and "$f_n$-circuit", and after the connection, "$f_m^{cnct}$-circuit" and "$f_n^{cnct}$-circuit". Even though the $f_m^{cnct}$-circuit and the $f_n^{cnct}$-circuit are multi-ports, they may be easily defined.

Using the $f$-connection, we compose a polynomial (in the sense of the form of $f(.)$) circuit from some $\alpha$-circuits. Interpretation of a polynomial circuit as such an "$\alpha$-connection", helpful in estimating the input current, was named the $\alpha$-test.

Work [1] described an *essentially experimental*, -- even when found via an analytical solution of circuit equations, -- *fact* that the input conductivity characteristic $F(.)$ (that is $i_{in} = F(v_{in})$) is very close to a linear form of the same powers of the independent variable as $f(.)$. This is named in [1] *approximate analytical superposition* in the sense of the map $f \to F$.

In the general case, the nodal voltages (besides the fixed $v_\mathbf{a} = v_{in}$ and $v_\mathbf{b} = 0$) are changed with the $f$-connection of two $\alpha$-circuits. Exceptional cases of ideal superposition when the nodal voltages are unchanged are discussed in [1]. The present part of the work reveals and carefully considers the reason for the good precision of the analytical superposition, observed in the regular cases, not associated with any special topology.

The structure of the present part of the work is as follows.

Section 2 recalls the problem of approximation of the current of the $\alpha$-connection $F(.)$ by the sum of the independent currents of the involved $\alpha$-circuits.

Section 3 explains why the value of $D$ in (1) may be, in many cases, taken as 1.

Section 4 gives a simple argument for the here basic assumption that in each $\alpha$-circuit, each of the $v_{s''}(\alpha)$ is a monotonic function, either increasing or decreasing.

Section 5 starts treating the *reason* for the high precision of the analytical superposition. In this section, the proof relates to the case when all of the $v_{s''}(\alpha)$ are similarly monotonic. Subsection 5.1 presents simulation results that well confirm the theory. Subsection 5.2 develops a theoretical boundary for the error of the superposition. In the latter development we simply assume that for the $\alpha$- (power-law) realization of the digraph, $v_{s''}(\alpha)$ are either known analytically, or found (say, by using a computer) for the particular $\{\alpha\}$ involved. The principle to use that the $\alpha$-realization of the digraph is relatively easy to describe lies at the foundation of the whole research, and it will be also relevant to all other boundaries of the error, which will be found in the sequel of the work, and this use of the relatively very simple $\alpha$-realization, gives the whole research a logical completion, and the $\alpha$-circuits seem to be a good theoretical tool.

Section 6 considers the case when not all of the $v_{s''}(\alpha)$ are similarly monotonic. This case includes the previous one as a particular.

Section 7 starts using power relations (energy conservation and the general Tellegen's theorem), in order to obtain other boundaries for the error of the superposition. We now use all of the nodal voltages of the circuit, calling this a transfer from the $\{s''\}$-representation of $F$ and $G$ to the $\{s\}$-representation. The preservation of the topology during *f*-connection makes application of the remarkable (but rarely used) Tellegen's theorem be natural here.

Section 8 briefly considers how the case of a two-term polynomial *f*(.) may be helpful in the analysis of cases of *f*(.) including three or four degrees. The arguments employ the possibility of obtaining *f*-connection by steps.

The numeration of the "Sentences" in [1] and here is common. We thus shall start in Section 5 from Sentence 2.

Recalling the main notations, terminology and definitions, introduced in [1]:

*f*(.) (or $i = f(v)$) – conductivity function of the similar elements involved in a realization of the given input topology. Work [1] introduced also resistive formulation of the circuit.

$i_{in} = F(v_{in})$ input current (input conductivity function) of the 1-port. Here $v_{in}$ is given. For the resistive formulation of the circuit, $i_{in}$ denotes the input current source that defines $v_{in}$.

*k* – integers labeling nodes.

*s* – integers labeling internal branches.

$\{s''\}$ – branches connected to the grounded input node that is always denoted as **b**; each such branch connects a respective node $k_{s''}$ to **b**; these branches are used for writing $i_{in} = F(v_{in})$. See also the **Appendix**.

$k_{s''}$ – the index for nodes that are close to the grounded input node **b**.

$w_{s''}$ – the number of the *parallel* branches connecting $k_{s''}$ with **b**; as a rule, $w_{s''} = 1$ in the present theory.

"*f*-circuit" – 1-port composed of conductors *f*(.). The 1-port may be of any topology. In view of a specific composition here of the 1-port from some other 1-ports of the same topology, it is useful to approach the "*f*-circuit" as a *"realization" of the given topology*.

$\varphi(\alpha)$ -- function defined by the equality $F_\alpha(v_{in}) = D\varphi(\alpha)(v_{in})^\alpha$. Such a function exists for any "$\alpha$-circuit".

$D$, $\alpha$ – positive constant parameters included in the specific case of $f_\alpha(.) = D(.)^\alpha$, when the 1-port is named "$\alpha$-*circuit*". $D$ appears to be a linear scaling parameter for all of the currents, including $i_{in}$; the role of $\alpha$ is much more interesting. For "$\alpha$-circuit", or "$f_\alpha$-circuit", $i_{in}$ is denoted as $F_\alpha(v_{in})$.

*$f_\alpha$-circuit* -- the same as "$\alpha$-circuit". For $\alpha$ integer, we also use the notation $f_p$-circuit; $p = 1,2, \ldots$ **P**. Such circuits are the "building blocks" that create the "$f$-circuit" in the sense of the (following) definition of "*f-connection*". The nodal voltages in an $\alpha$-circuit are denoted as $v_k(\alpha)$.

*"f-connection"* -- the node-to-node connection of two "$f$-circuits", keeping the topology of the 1-port, as is demonstrated by Fig.2. From the analytical side, "$f$-connection" just changes $f(.)$, or the *realization* of the topology (digraph).

$f_\alpha^{cnct}$-*circuit* (*or $f_p^{cnct}$-circuit, for $\alpha$ integer*) – the *part* (*generally, a multi-port*) of an $f$-circuit, which originates from the particular $f_\alpha$-circuit (or $f_p$-circuit) involved in the creation of the "$f$-circuit" by means of the "$f$-connection". (Consider each of the two "wings" of the circuit of Fig. 2 *after* all the respective nodes are connected.) That the $f_\alpha^{cnct}$-circuit is, contrary to the $f_\alpha$-circuit, not a 1-port means that "$f$-connection" of the 1-ports is a "destroying" connection. However, there is no difficulty of seeing an "$f_\alpha^{cnct}$-circuit" in the $f$-connection. The input-current function of this $f_\alpha^{cnct}$-circuit is denoted as $F_\alpha^{cnct}$. It is obvious that for the $f$-connection $F(.) = (\Sigma F_\alpha^{cnct})(.)$.

"*Approximate analytical (structural) superposition*" -- approximation of $F(.) = (\Sigma F_\alpha^{cnct})(.)$ by means of $G(.) = (\Sigma F_\alpha)(.)$.

It is proved that the main (of the smallest degree) terms of $F(.)$ and $G(.)$ are the same, and it is also proved that for some specific topologies, $G(.) \equiv F(.)$ *absolutely precisely*, i.e. the superposition is ideal. In this case, the nodal potentials $v_k$ necessarily are independent of $\alpha$, i.e. are the same for any "$f_\alpha(.)$-realization", and for the $f(.)$-connection. In this case, the $f_\alpha^{cnct}$-circuits *actually* remain the same initially given 1-ports "$f_\alpha$-circuits", and the "$f$-connection" is equivalent to the usual parallel connection of the given $f_\alpha$-circuits, when $i_{in} = (\Sigma F_\alpha)(.)$. Such specific topologies are, however, very rare.

Another simple case is that when *all* of the $\alpha$ involved are large (actually $\alpha \geq 3$ is already the "large" value). The superposition is then necessarily very good, for any topology. Such cases of nonlinearity are, however, also a very rare case.

$\eta = |F-G|/F$ – relative error of the approximation of $F$ by $G$. According to *Statement 1* of [1] (saying that $\lim F(x)/G(x) = 1$, as $x \to 0$), for any $f$-connection, $\eta \to 0$ as $v_{in} \to 0$, and for a linear circuit, $\eta = 0$

$\alpha$-*test* – interpretation of $f$-circuit, with '$f$' composed of several (here usually two) degrees as $f$-connection of some $\alpha$-circuits.

## 2. The "internal circuit mechanism" of the approximation of $F(.)$ by $G(.)$ for 1-ports with polynomial (monotonic) $f(.)$

Speaking about a general case of $f$-circuit, we consider that though the dependence of the nodal potentials on $\alpha$, and the consequent change of $v_k$ with the $\alpha$-connection, prevents ideal superposition from being obtained, the dependence $v_k(\alpha)$ features a "regulation" of the potentials, which leads to the relatively high precision of the superposition.

We use below integers $\{\alpha_1,\alpha_2\} = \{m,n\}$ (thus it is suitable to refer to Fig. 2), but it is obvious that the following conclusions relate to any $\{\alpha_1,\alpha_2\}$.

The nodal voltages in the two separated 1-ports are denoted, respectively, as $v_k(m)$ and $v_k(n)$, and the nodal voltages in the $f$-connection simply as $v_k$. As in [1], the input currents of the "$m$" and "$n$" circuits, taken separately, will be denoted as $F_m(.)$ and $F_n(.)$, and in the *f*-connection as $F_m^{cnct}(.)$ and $F_n^{cnct}(.)$, respectively, and the total input current of the *f*-connection as $F(.)$.

Branches $s''$ are those directly connected to the grounded node **b**, and no two such are connected in parallel (i.e. $w_{s''} = 1$, $\forall s''$). In these notations (see also the Appendix), for the *f*-connection, KCL *at the grounded input node* **b** is

$$F(v_{in}) \equiv F_m^{cnct}(v_{in}) + F_n^{cnct}(v_{in}) = \sum_{s''}[D_m v_{s''}^m + D_n v_{s''}^n]. \qquad (2)$$

According to the conception of analytical superposition [1], expression (2) has to be compared with

$$G(v_{in}) \equiv F_m(v_{in}) + F_n(v_{in}) = \sum_{s''}[D_m v_{s''}^m(m) + D_n v_{s''}^n(n)] \qquad (3)$$

including the input currents of the *separated* (*connected in parallel*) '*m*' and '*n*' circuits.

Since the node **b** *collects all* the internal currents in each branch $s''$, the respective elements of the incident matrix are the same for these currents. We took these elements in (2,3) as +1; the factor -1 would be contracted, anyway.

We precede the proofs by some simple, but very important comments.

## 3. Comments regarding comparison of the different realizations {*D*,α} of the *f*-circuit

We shall compare the realizations with $f_m(.) = D_m(.)^m$ and $f_n(.) = D_n(.)^n$. Clearly, only for $D_m = D_n$ changing $\alpha$ from *m* to *n* is completely equivalent to passing on from one of the circuits to the other one. However the comparison of the sets $v_{s''}(m)$ and $v_{s''}(n)$ does not require $D_m = D_n$. No such limitation exists since the value of *D* in (1) does not influence any $v_k$, and since only the values of the nodal potentials are important to the point, we can change $D_m$ to $D_n$ together with the change of *m* to *n*. In other words, in the context of the study of $v_{s''}(\alpha)$, it is the same thing to just change $\alpha$, or the whole (*D*,α)-realization of the digraph.

To illustrate the point, consider the role of the inequality ($m < n$, and $v_m < v_n$)

$$(D_m + D_n)v^m(m) < D_m v^m + D_n v^n < (D_m + D_n)v^n(n)$$

in an analysis that compares the values of $\{v_k\}$ for the case when $f(v) = D_m v^m + D_n v^n$ with the respective values $v(m)$ and $v(n)$ for the case of a single-degree $f(.)$. One concludes that not $D_m$ and $D_n$, but just the degrees are important.

Coefficients $D_m$ and $D_n$ may be interesting (Section 8.3) for just observing how $D_m v^m + D_n v^n$ is converted to a single degree expression when one of the coefficients tends to zero. Thus, in the following derivations we sometimes set for simplicity $D_m = D_n = 1$. There is, however, a problem where it is necessary to change *D* together with α. In extended work [3], we write $f(.)$ not only in the simplified writing (1) in which $D_m$ and $D_n$ have different physical dimensions, but also as $i = i_o(v/v_o)^\alpha$ with two constants, $i_o$ and $v_o$ having, respectively, the dimensions of current and voltage. In comparison with such notation, $D = D(\alpha) = i_o v_o^{-\alpha}$, and a change in α changes $D$. However such a detailed representation of $f_\alpha(.)$ appears to be necessary only for a detailed study in [3] of the transfer to voltage hardlimiters, as $\alpha \to \infty$.

## 4. Monotonicity of $v_{s''}(\alpha)$ and its role for the analytical superposition

<u>Observation</u>:  Besides $\alpha$, no non-dimensional parameter is given in the formulation of the "$\alpha$-circuit", which could be compared with $\alpha$ and give to a value of $\alpha$ some specific analytical role, in any possible formula.

<u>Consequence 4-1</u>: The functions $v_k(\alpha)$ may not have extremes at any value of $\alpha$. In particular, this relates to the nodes with $k \in \{k_{s''}: v_{s''} = v_{k_{s''}} - v_b = v_{k_{s''}}\}$ (see the Appendix). We thus find that all of the $v_{s''}(\alpha)$, excluding the possibly present input **ab**-conductor with $v \equiv v_{in}$, are some *monotonic* functions of $\alpha$, either increasing or decreasing.  This must result in the double inequality for the $\{v_{s''}\}$ of the $f$-connection.

$$min\{v_{s''}(m), v_{s''}(n)\} < v_{s''} < max\{v_{s''}(m), v_{s''}(n)\}, \quad \forall s''. \qquad (4)$$

<u>Consequence 4-2</u> :   For any $s''$,  $v_{s''}$ cannot equal only one of the values $\{v_{s''}(m), v_{s''}(n)\}$ ($m \neq n$). Indeed, if  $v_{s''} = v_{s''}(m) \neq v_{s''}(n)$,  then  $\alpha = $'$m$' must be some analytically specified value, but no such value exists. The equalities  $v_{s''} = v_{s''}(m) = v_{s''}(n)$  may take place only simultaneously, which is the specific case of the ideal superposition relevant only to special topologies, as discussed in Section 4 of [1]. We shall often ignore this trivial possibility below.  □

For many circuits, all of $v_{s''}(\alpha)$ are *similarly* monotonic, i.e. all either increasing or decreasing, which means that *for two $\alpha$-circuits of the same topology, one with $\alpha = m$, and another with $\alpha = n$, $m \neq n$, all the $v_{s''}(m)$ (but not all $v_s$ [1]) are either larger or smaller than the respective $v_{s''}(n)$*. We start (Statement 2) the proofs from this case for ease of writing, and numerical example in Section 5.1 also relates to this case. (See also examples of Sections 6 and 7 in [1].)
In this case

$$sign\frac{dv_{s''}(\alpha)}{d\alpha} = const, \quad \forall \ s''.$$

However, Appendix 1 of [1] also gives an example of a circuit for which

$$sign\frac{dv_{s''}(\alpha)}{d\alpha} = \pm 1,$$

depending on what $s''$ is. This simply means that  for some $s''$  $v_{s''}(m) < v_{s''}(n)$, while for the other $s''$  $v_{s''}(n) < v_{s''}(m)$.
This case, which will be treated in Statement 3, obviously includes the previous case as a particular one.

## 5. The case when all of the $v_{s''}(\alpha)$ are similarly monotonic

Statement 2:

$$|F(v_{in}) - G(v_{in})| < max\{|F_m^{cnct}(v_{in}) - F_m(v_{in})|, |F_n^{cnct}(v_{in}) - F_n(v_{in})|\}. \quad (5)$$

*That is, with the α-connection, the sum of the input currents of the 'm' and 'n' circuits is changed more weakly (and since $F > max\{F_m, F_n\}$, <u>relatively</u> much more weakly) than is at least one of the currents.* □

The reason for (4) to be correct is that if the individual input currents of the circuits are changed with the "*f*-connection" of these circuits, they are changed *oppositely*, with one increased, and one decreased. That is,

$$sign[F_m(v_{in}) - F_m^{cnct}(v_{in})] = - sign[F_n(v_{in}) - F_n^{cnct}(v_{in})]$$

and

$$|F(v_{in}) - G(v_{in})| \equiv |F_m^{cnct}(v_{in}) + F_n^{cnct}(v_{in}) - (F_m(v_{in}) + F_n(v_{in}))| \equiv$$

$$\equiv |(F_m^{cnct}(v_{in}) - F_m(v_{in})) + (F_n^{cnct}(v_{in}) - F_n(v_{in}))|. \quad \square \quad (6)$$

Proof: Let us assume, for certainty, that $v_{s''}(m) < v_{s''}(n)$. This means that (4) may be rewritten as

$$v_{s''}(m) < v_{s''} < v_{s''}(n), \quad \forall s''. \quad (7)$$

Compare the right-hand sides of the expressions

$$F_m^{cnct}(v_{in}) - F_m(v_{in}) = \sum_{s''} D_m[v_{s''}^m - v_{s''}^m(m)]$$

and

$$F_n^{cnct}(v_{in}) - F_n(v_{in}) = \sum_{s''} D_n[v_{s''}^n - v_{s''}^n(n)]$$

included in (6). Since it follows from (7) that

$$v_{s''}^m > v_{s''}^m(m), \quad \forall s''$$

and

$$v_{s''}^n < v_{s''}^n(n), \quad \forall s'',$$

the above sums have different signs, which in view of (6) proves (5). □

Remark 5-1: Observe that the numerical simulations reported in Section 5.1 show that for two α-circuits with $D_1 = D_2$ and $\alpha_1 = 1$, $\alpha_2 = 3$, *the (dominant) input current of the circuit with lower α increases* with the α-connection, *and* the (small) *current of the circuit with higher $\alpha$ decreases.* Observe also that for the circuit with higher $\alpha$ the *relative change* in the current is stronger and the total current is slightly *decreased*, i.e. in this case $F < G$. Since it follows from the latter inequality that $v_{in}F(v_{in}) < v_{in}G(v_{in}) = v_{in}F_m(v_{in}) + v_{in}F_n(v_{in})$, the total power consumption by $\alpha$-connecting may be decreased in some cases. □

### *5.1. A MatLab circuit for the "internal" study of the analytical superposition*

The circuit shown in Fig. 3 has been studied using MatLab. We illustrate the above conclusions by measurements of $F_m(v_{in})$, $F_n(v_{in})$, $F_m^{cnct}(v_{in})$ and $F_n^{cnct}(v_{in})$; for the comparison of $F_m(v_{in})$ with $F_m^{cnct}(v_{in})$, $F_n(v_{in})$ with $F_n^{cnct}(v_{in})$, and $F_m^{cnct}(v_{in}) + F_n^{cnct}(v_{in})$ with $F_m(v_{in}) + F_n(v_{in})$ (i.e. $F(v_{in})$ with $G(v_{in})$). We set $m = 1$ and $n = 3$.

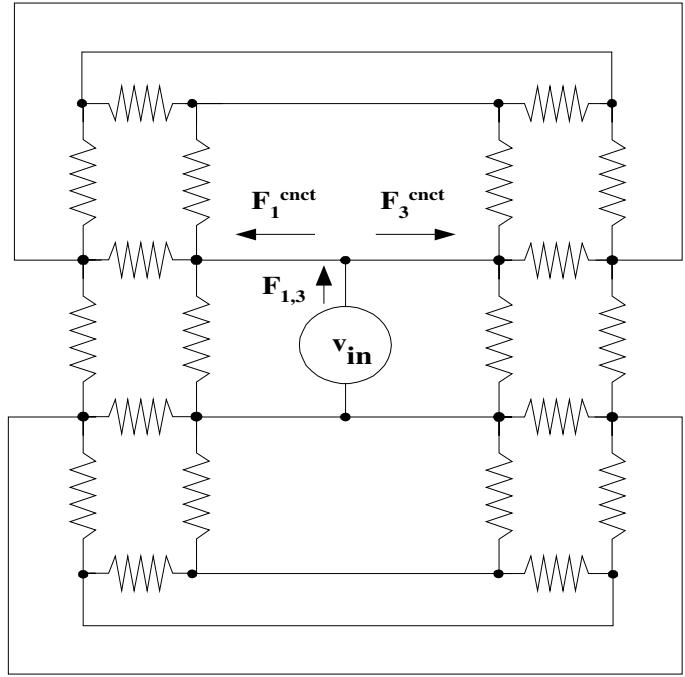

Fig. 3: The "$\alpha$-connection" for directly checking the 'analytical superposition'. In the left 'wing', $\alpha = 1$, and in the right $\alpha = 3$. $D_1 = D_3 = 1$. We measure the particular input currents before, and after connection, and also the total current $F_{1,3}$ (denoted in the above formulae as '$F$') in the connected state. $v_{in}$ is taken as 1V.

The left and the right 'wings' of the circuit are the '$\alpha$-connected' 1-ports. For the left, $\alpha = 1$, and for the right $\alpha = 3$. We set $v_{in} = 1$V.

The results are collected in Table 1.

| The circuit | $F_{(\alpha)}$ | $F_{(\alpha)}^{cnct}$ | | Error in the analytical superposition. |
|---|---|---|---|---|
| $\alpha = 1$ | 1.4 | 1.466 | +4.7% | |
| $\alpha = 3$ | 1.14 | 1.044 | -8.4% | |
| *f*-connection ($F_{1,3}$, or $F$) | 2.511 | | | |
| Parallel connection ($G$) | 2.54 | | | **1.15%** |

Table 1: The results of the MatLab simulation for the circuit of Fig. 3.

We see that input currents $F_1$ and $F_2$, of the wings *taken separately* (before the connection), were measured as 1.4 A for $\alpha = 1$, and 1.14 A for $\alpha = 3$. That is, $G = 2.54$A. For the *f*-connection $i_{in} = F = 2.511$A was measured. The relative error in the analytical superposition here is thus $(F - G)/F = (2.54 - 2.511)/2.511 = 0.0115$, i.e. 1.15%.

We also separately measure the input currents $F_1^{cnct}$ and $F_3^{cnct}$ of the 'wings', in the *f*-connected state. The values $F_1^{cnct} = 1.466$ A, and $F_3^{cnct} = 1.044$ A were obtained. That is, *in the f-connected state the input current of the less nonlinear circuit ($\alpha = 1$) increases* (from 1.4 to 1.466), *and that of the more nonlinear circuit ($\alpha = 3$) decreases* (from 1.14 to 1.044). The associated relative changes in the wing's currents are +4.7% and -8.2% (the strongest change is for the weakest current), this to be compared with the change of 1.15% for the total current.

When passing from the usual parallel to the *f*-connection, the total input current is changed much more weakly than the particular input currents.

Since the particular currents may be changed only because of the change in the internal nodal voltages, it is obvious from the above data that the nodal voltages also changed relatively strongly (several percent) with the *f*-connection. That is, the circuit is certainly far from the conditions that ensure (Section 4 of [1]) ideal superposition.

### *5.2. A boundary for the error of the analytical superposition*

Since it is much simpler to precisely calculate the $\alpha$-circuit than the *f*-connection, determining $\{v_s(m)\}$ and $\{v_s(n)\}$, we shall obtain a boundary for the relative error $\eta = |F - G|/F$ by comparison of a sum including $\{v_s\}$ with some sums that include $\{v_s(m)\}$ and $\{v_s(n)\}$, assuming that the latter parameters are known.

Regarding the basic equality

$$F - G = \sum_{s''}[v_{s''}^m + v_{s''}^n - v_{s''}^m(m) - v_{s''}^n(n)]$$

consider two in principle possible cases; firstly $F > G$, when

$$\eta F = |F - G| = F - G = \sum_{s''}[v_{s''}^m + v_{s''}^n - v_{s''}^m(m) - v_{s''}^n(n)] \;, \quad (8)$$

and then $G > F$, when

$$\eta F = |F - G| = G - F = \sum_{s''}[v_{s''}^m(m) + v_{s''}^n(n) - v_{s''}^m - v_{s''}^n]. \quad (9)$$

In the case of (8), we use the right inequality in

$$v_{s''}(m) < v_{s''} < v_{s''}(n), \quad \forall s'', \qquad (7) \text{ (repeated)}$$

and increase the right-hand side of (8) by substituting $v_{s''}(n)$ instead of $v_{s''}$. We thus obtain (note that the terms $\pm v_{s''}^n(n)$ are canceled) the following inequality:

$$|F - G| < \sum_{s''}[v_{s''}^m(n) - v_{s''}^m(m)].$$

In the case of (9), we use the *left* side of (7), and by very similar operations (substituting $v_{s''}(m)$ instead of $v_{s''}$, etc.) obtain:

$$|F - G| < \sum_{s''}[v_{s''}^n(m) - v_{s''}^n(n)]$$

Using the concluding inequality for $|F - G|$

$$|F - G| < \max\left\{\sum_{s''}[v_{s''}^m(n) - v_{s''}^m(m)]; \sum_{s''}[v_{s''}^n(m) - v_{s''}^n(n)]\right\}, \quad (10)$$

and an upper boundary for the relative error, we can now write

$$\eta = \frac{|F - G|}{F} < \frac{\max\left\{\sum_{s''}[v_{s''}^m(n) - v_{s''}^m(m)]; \sum_{s''}[v_{s''}^n(m) - v_{s''}^n(n)]\right\}}{\sum_{s''}[v_{s''}^m(m) + v_{s''}^n(m)]} \quad (11)$$

where we reduced $F$ in the denominator, by replacing $v_{s''}$ by $v_{s''}(m)$, in accordance with (7).

The right-hand side of (11) is expressed via the relevant values of the functions $\{v_{s''}(\alpha)\}$ that can be found using the $\alpha$-test of the *f*-circuit. (The sets of *numerical values* $v_{s''}(m)$ and $v_{s''}(n)$ also may be directly found using computer.)

Noting that

$$\eta = \frac{|F-G|}{F} = \frac{|F-G|}{G-(G-F)} \le \frac{|F-G|}{G-|G-F|} = \frac{|F-G|}{G-|F-G|}, \qquad (12)$$

we can substitute into (12) the upper boundary for $|P_F - P_G|$, which gives

$$\eta < \frac{\max\left\{\sum_{s''}[v_{s''}^m(n) - v_{s''}^m(m)]; \sum_{s''}[v_{s''}^n(m) - v_{s''}^n(n)]\right\}}{G - \max\left\{\sum_{s''}[v_{s''}^m(n) - v_{s''}^m(m)]; \sum_{s''}[v_{s''}^n(m) - v_{s''}^n(n)]\right\}}.$$

In Section 7 we shall find some other boundaries, expressed not via $\{v_{s''}(\alpha)\}$, but via *all of the* $\{v_s(\alpha)\}$. However the principle of obtaining any such boundary by being based on the solution of the relevant $\alpha$-circuit(s), will remain.

## 6. The case when $v_{s''}(\alpha)$ are differently monotonic

For some circuits, $dv_{s''}(\alpha)/d\alpha$ may be of different signs. That is, for some nodes close to **b**, $v_{s''}(m) < v_{s''}(n)$, and for some $v_{s''}(n) < v_{s''}(m)$. Figure A1-2 of [1] gives example of such circuit. Let us return to this example (see Fig. 4), discussing it more deeply.

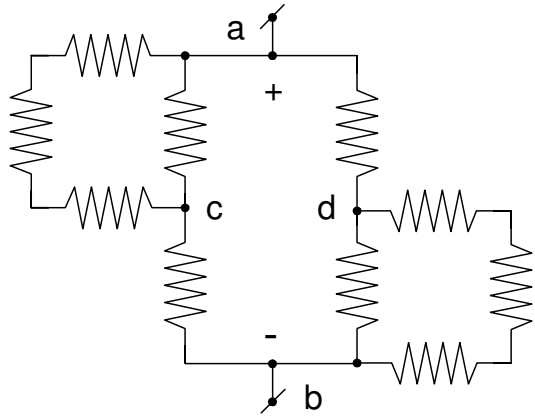

Fig. 4: Fig. A1-2 of [1] repeated for a deeper discussion. This time, we do not focus on the symmetry. The symmetry and the parallel-type composition may be eliminated, while keeping the different

sign[$dv_{s''}(\alpha)/d\alpha$], dependent on $s''$. By the way, this example shows that the fact that $dv_{s''}(\alpha)/d\alpha$ are of different signs, need not spoil or improve the analytical superposition.

In this circuit, sign[$dv_c(\alpha)/d\alpha$] = - sign[$dv_d(\alpha)/d\alpha$], i.e. $v_{s''}(\alpha)$, though being necessarily both monotonic, behave non-similarly in the sense of decrease or increase.

A possible (not as in [1]) argument here may be that with increase in $\alpha$, when the conductors become similar voltage hardlimiters, the three series conductors become blocked by one conductor, in both sides. Thus, the voltages **ac** and **db** are both increased in the voltage division. This means that $v_c$ becomes closer to $v_b = 0$, i.e. decreasing, while $v_d$ becomes increasing. □

One can, of course interpret this circuit as a parallel connection of two circuits, each treatable according to Statement 2, and $F(.)$ of the circuit is composed of two (equal) terms for each of which the approximate analytical superposition works as before. However, for a more complicated circuit such a simple decomposition as in the above circuit may be impossible and the symmetry is also not necessary, but {$dv_{s''}(\alpha)/d\alpha$} may be of different sign. Introducing between the nodes **c** and **d** a high-ohmic connection, and adding to one of the sides one more conductor to those three that are already in series, we destroy both the composition and the symmetry, but sign[$dv_c(\alpha)/d\alpha$] = - sign[$dv_d(\alpha)/d\alpha$] may remain.

<u>Statement 3</u>: *Also in the general case, $F(v_{in}) - G(v_{in})$ is composed of equal number of positive and negative terms, which decreases $|F(v_{in}) - G(v_{in})|$.* □

<u>Proof</u>: Let us present, in the general situation, $\{s''\} = \{s''\}_1 \cup \{s''\}_2$; where, by the definition of the sets, for $\{s''\}_1$:

$$v_{s''}(m) < v_{s''} < v_{s''}(n), \quad \forall s'' \in \{s''\}_1,$$

and for $\{s''\}_2$:

$$v_{s''}(n) < v_{s''} < v_{s''}(m), \quad \forall s'' \in \{s''\}_2.$$

Using $\{s''\}_1$ and $\{s''\}_2$, we write now the involved sums by $s''$ as two parts:

$$F_m^{cnct}(v_{in}) - F_m(v_{in}) = \sum_{\{s''\}_1} D_m[v_{s''}^m - v_{s''}^m(m)] + \sum_{\{s''\}_2} D_m[v_{s''}^m - v_{s''}^m(m)] \quad (13)$$

and

$$F_n^{cnct}(v_{in}) - F_n(v_{in}) = \sum_{\{s''\}_1} D_n[v_{s''}^n - v_{s''}^n(n)] + \sum_{\{s''\}_2} D_n[v_{s''}^n - v_{s''}^n(n)], \quad (14)$$

having sums of different polarities in both right-hand sides. Thus, summing and putting at the last step the two positive sums in the first place, we write

$$|F(v_{in}) - G(v_{in})| \equiv |F_m^{cnct}(v_{in}) - F_m(v_{in}) + (F_n^{cnct}(v_{in}) - F_n(v_{in}))| \equiv$$

$$|\sum_{\{s''\}_1} D_m[v_{s''}^m - v_{s''}^m(m)] + \sum_{\{s''\}_2} D_n[v_{s''}^n - v_{s''}^n(n)] + \sum_{\{s''\}_2} D_m[v_{s''}^m - v_{s''}^m(m)] + \sum_{\{s''\}_1} D_n[v_{s''}^n - v_{s''}^n(n)]|$$

□

Based on the above equality for $|F_m(v_{in}) - G(v_{in})|$, and using the fact of the different polarities of the two first and the two last sums, we can write:

$$|F(v_{in}) - G(v_{in})| <$$

$$< \max\{\sum_{\{s''\}_1} D_m[v_{s''}^m - v_{s''}^m(m)] + \sum_{\{s''\}_2} D_n[v_{s''}^n - v_{s''}^n(n)]; |\sum_{\{s''\}_2} D_m[v_{s''}^m - v_{s''}^m(m)] + \sum_{\{s''\}_1} D_n[v_{s''}^n - v_{s''}^n(n)]|\}$$

(15)

Inequality (15) is *not* of the form (5). If, however,

$$\sum_{\{s''\}_1} D_m[v_{s''}^m - v_{s''}^m(m)] + \sum_{\{s''\}_2} D_m[v_{s''}^m - v_{s''}^m(m)] \equiv \sum_{s''} D_m[v_{s''}^m - v_{s''}^m(m)] > 0$$

while

$$\sum_{\{s''\}_1} D_n[v_{s''}^n - v^n(n)] + \sum_{\{s''\}_2} D_n[v_{s''}^n - v_{s''}^n(n)] \equiv \sum_{s''} D_n[v_{s''}^n - v_{s''}^n(n)] < 0$$

(or conversely regarding >,<), then we can write:

$$|F(v_{in}) - G(v_{in})| < \max\{|F_m^{cnct}(v_{in}) - F_m(v_{in})|, |F_n^{cnct}(v_{in}) - F_n(v_{in})|\}.$$

This analogy to form (5) form can be a better estimation than (10) because each of the expressions

$$F_m^{cnct}(v_{in}) - F_m(v_{in})$$

and

$$|F_n^{cnct}(v_{in}) - F_n(v_{in})|$$

includes, in this case, a positive and a negative sum. One notes that the latter circumstance, provided by the different monotonicity of $dv_{s''}(\alpha)/d\alpha$ may, in principle, even improve the superposition with regard to the case when $\text{sign}[dv_{s''}(\alpha)/d\alpha] = \text{const}, \forall s''$. However, one has to be careful with such insistence, as the example of Fig. 4 shows.

It may be also possible to write

$$|F(v_{in}) - G(v_{in})| <$$

$$\max\{|\sum_{\{s''\}_1} D_m[v_{s''}^m - v_{s''}^m(m)] + \sum_{\{s''\}_1} D_n[v_{s''}^n - v_{s''}^n(n)]|; |\sum_{\{s''\}_2} D_m[v_{s''}^m - v_{s''}^m(m)] + \sum_{\{s''\}_2} D_n[v_{s''}^n - v_{s''}^n(n)]|\}$$

if the sums grouped inside the first and the second "||" are of alternative polarity.

### 7. Another boundary for the error of the analytical superposition

In order to obtain a boundary other than that in Section 5.2, we transfer to a representation of $F$, $G$ and $\eta$ via *all* of the $v_s$ (or $v_s$). Again, the separate $\alpha$-circuits involved are much more simply calculated than the *f*-connection, and we shall estimate $\eta$ by $\{v_s(m)\}$ and $\{v_s(n)\}$. One may also consider that any full circuit description is given, in principle, by a system of equations for all of the $v_s$, and that because of the nonlinearity of this system it may be difficult, in the general case, to separately obtain $\{v_{s''}\}$. Thus, the idea of the obtaining the alternative boundary is to compare a sum including $\{v_s\}$ with some sums, including either $\{v_s(m)\}$ or $\{v_s(n)\}$, assuming that the latter variables are known.

<u>Definition 1</u>: From now on, our previous representation of *F* or *G* via $\{v_{s''}\}$ will be named its $\{s''\}$-*representation*, and the new representation via all of the $\{v_s\}$ will be named $\{s\}$-*representation*. □

The transfer to the $\{s\}$- representation will be done using energy conservation, which introduces all of the $\{v_s\}$ via $\{v_s i_s\}$. We shall not need to consider whether or not certain monotonicity of $v_k(\alpha)$ yields certain monotonicity of $v_s(\alpha)$; the following consideration includes the cases of different $\text{sign}[dv_s(\alpha)/d\alpha]$.

### *7.1. The transfer to the $\{s\}$-representation and a boundary for $\eta$*

Because of energy conservation, the input power of the *f*-circuit equals the power dissipated in all of the elements of this purely resistive circuit ($P_F$ denotes the power of the *f*-connection and $P_G$ of the parallel connection, and for simplicity of writing, we set below $D_m = D_n = 1$.):

$$P_F = v_{in} F(v_{in}) = \sum_s v_s i_s = \sum_s v_s (v_s^m + v_s^n) = \sum_s (v_s^{m+1} + v_s^{n+1}) , \qquad (16)$$

when summing over all internal branches.
That is

$$F(v_{in}) = \sum_s \frac{v_s}{v_{in}} (v_s^m + v_s^n) = \frac{1}{v_{in}} \sum_s (v_s^{m+1} + v_s^{n+1}), \qquad (17)$$

and for the parallel connection of the *m* and *n* circuits:

$$P_G = v_{in} G(v_{in}) = \sum_s v_s(m) i_s(m) + \sum_s v_s(n) i_s(n) = \sum_s [v_s^{m+1}(m) + v_s^{n+1}(n)]$$

i.e.

$$G(v_{in}) = \frac{P_G}{v_{in}} = \frac{1}{v_{in}} \sum_s (v_s^{m+1} + v_s^{n+1}). \qquad (18)$$

from (16) and (18)

$$\eta = \frac{|F - G|}{F} = \frac{|P_F - P_G|}{P_F}, \qquad (19)$$

or, for the *non-relative positive* error's measure, $\eta F$,

$$\eta F = \frac{|P_F - P_G|}{v_{in}} = |F - G|, \tag{20}$$

which also is a useful variable. In accordance with the above, we assume that $G$ and $P_G$ are known.

Having an estimation for the non-relative error, one can bound $\eta$. Indeed,

$$\eta = \frac{|P_F - P_G|}{P_F} = \frac{|P_F - P_G|}{P_G - (P_G - P_F)} \leq \frac{|P_F - P_G|}{P_G - |P_G - P_F|} = \frac{|P_F - P_G|}{P_G - |P_F - P_G|}, \tag{21}$$

where $P_G$ is known.

We thus shall proceed not with $\eta$, but with the simpler "non-relative error" $|P_F - P_G|$ = $v_{in}|F - G|$.

Using (20), we obtain:

$$\eta v_{in} = \left| \sum_s [v_s^{m+1} + v_s^{n+1} - v_s^{m+1}(m) - v_s^{n+1}(n)] \right|, \tag{22}$$

which could be also written as

$$\frac{|\Delta P_m + \Delta P_n|}{v_{in}}$$

where

$$\Delta P_m = \sum_s [v_s^{m+1} - v_s^{m+1}(m)] \quad \text{and} \quad \Delta P_n = \sum_s [v_s^{n+1} - v_s^{n+1}(n)]$$

are the changes in the power consumption of the $m$ and $n$ circuits (the power in the connected state minus the power in the non-connected state). This corresponds to writing $\eta$ as

$$\eta = \frac{|\Delta P_m + \Delta P_n|}{P_F}, \tag{22a}$$

and the results of the previous sections suggest that $\text{sign}[\Delta P_m] = - \text{sign}[\Delta P_n]$. However we shall proceed with the form (22).

Remark 7-1: The ease of passing in the above formulae, from the current functions to the powers and conversely, should not hide the fact that we came to the $\{s\}$-representation in *F via the powers*. □

In order to obtain a boundary for the error, we consider that since $v_s(m)$ and $v_s(n)$ are assumed to be known from solution of the $\alpha$-circuit, it is not a problem, in principle, to divide $\{s\}$ into two groups: $\{s\}_1$ for which

$$v_s(m) < v_s(n)$$

i.e.

$$v_s(m) < v_s < v_s(n) \tag{23}$$

and $\{s\}_2$ for which

$$v_s(n) < v_s(m)$$

i.e.

$$v_s(n) < v_s < v_s(m) . \qquad (24)$$

Correspondingly, we use for set $\{s\}_1$ that $v_s < v_s(n)$, and obtain (by simply substituting $v_s = v_s(n)$, which leads, in particular, to contraction of two terms) the following inequality:

$$v_s^{m+1} + v_s^{n+1} - v_s^{m+1}(m) - v_s^{n+1}(n) < v_s^{m+1}(n) - v_s^{m+1}(m) , \quad s \in \{s\}_1,$$

and for $\{s\}_2$, using that $v_s < v_s(m)$, we similarly obtain

$$v_s^{m+1} + v_s^{n+1} - v_s^{m+1}(m) - v_s^{n+1}(n) < v_s^{n+1}(m) - v_s^{n+1}(n) , \quad s \in \{s\}_1 .$$

Thus, we have

$$\eta F(v_{in}) < \sum_{\{s\}_1} [v_s^{m+1}(n) - v_s^{m+1}(m)] + \sum_{\{s\}_2} [v_s^{n+1}(m) - v_s^{n+1}(n)] . \qquad (25)$$

If $F > G$, then the sign of the absolute value in (22) can be omitted, and one sees that (25) gives the upper boundary for $\eta F$, and, according to (16a), also for $\eta F$.

If, however, $G > F$, then

$$|F - G| = G - F \sim P_G - P_F ,$$

and we have to find a boundary for the positive sum

$$\sum_s [v_s^{m+1}(m) + v_s^{n+1}(n) - v_s^{m+1} - v_s^{n+1}] .$$

In this case, after separating $\{s\}$ into $\{s\}_1$ and $\{s\}_2$, we now use the *left sides* of (23,24), i.e.

$$v_s > v_s(m) \qquad s \in \{s\}_1$$

and

$$v_s > v_s(n) \qquad s \in \{s\}_2,$$

and by replacing $v_s$ by the smaller value (respective for each sum), immediately obtain

$$\eta F < \sum_{\{s\}_1} [v_s^{n+1}(n) - v_s^{n+1}(m)] + \sum_{\{s\}_2} [v_s^{m+1}(m) - v_s^{m+1}(n)] \qquad (26)$$

(observe that after mutual replacement/interchange of the signs $\{s\}_1$ and $\{s\}_2$ under the sums in (26), the latter boundary becomes the negative of the boundary in (25)).

From both of the above cases, we have, finally, the general boundary as

$$\eta F < \{\text{maximum of the boundaries (25,26)}\}.$$

Since, however, for all of the concrete circuits studied it was found that $G > F$, boundary (26) can be assumed to be sufficient.

It is easy now to proceed in either of the above ways (e.g. that which led to (10)), obtaining another new boundary for $\eta$. However, it is more important to demonstrate that the $\{s\}$-representation and the invariance of the topology allow us to obtain two other boundaries, using Tellegen's theorem.

*7.2 Application of Tellegen's theorem to the boundary estimation*

The similar topology of all of the circuits involved in these considerations suggests that we to try to improve the estimation of the boundary, using Tellegen's theorem [4-7]:

$$\mathbf{v}^{(1)\text{T}}\mathbf{i}^{(2)} = -v_{in}^{(1)}i_{in}^{(2)} + \sum_s v_s^{(1)} i_s^{(2)} = 0 \qquad (27)$$

or

$$v_{in}^{(1)} i_{in}^{(2)} = \sum_s v_s^{(1)} i_s^{(2)} ,$$

where $\mathbf{v}^{(1)}$ is the vector of $\{v_s^{(1)}\}$, for one realization of the digraph, and $\mathbf{i}^{(2)}$ the vector of $\{i_s^{(2)}\}$ for (generally) another realization. The input-terminals term is written here separately, because the index '$s$' labels only internal branches of the circuit. Sign '-' in [22] is because $i_{in}$ comes out of the +'ve terminal of the input source.

The input voltage is the same, while the input current may here be $F(v_{in})$, $G(v_{in})$, $G_m(v_{in})$ and $G_n(v_{in})$.

Thus, we have, by taking in the energy relation (27) voltages in the connected state ("$F$-circuit") and the currents of the separate $m$-circuit, that

$$v_{in} G_m(v_{in}) = \sum_s v_s v_s^m(m) ,$$

and similarly, using the currents of the separate $n$-circuit

$$v_{in} G_n(v_{in}) = \sum_s v_s v_s^n(n) .$$

Adding these equations we obtain

$$v_{in} G(v_{in}) = \sum_s v_s [v_s^m(m) + v_s^n(n)] . \qquad (28)$$

We now obtain by subtraction of (28) from (16) that

$$v_{in} \cdot (F-G)(v_{in}) = \sum_s v_s [v_s^m - v_s^m(m) + v_s^n - v_s^n(n)]. \qquad (29)$$

As in Section 6, we can present $\{s\} = \{s\}_1 \cup \{s\}_2$, and replace $v_s$ by $v_s(m)$, or $v_s(n)$, in the $\{s\}_1$-sum, or $\{s\}_2$-sum, which leads to boundaries that are somewhat different from (25, 26). We shall not compare here differently obtained boundaries for the error.

It has to be stressed that since the "*f*-connection" involves only the circuits with the same digraphs, we have here a very suitable field for application of the Tellegen's theorem.

## 8. On the $\alpha$-test of *f*-circuits with third and fourth degree polynomial *f*(.)

The cases of *f*(.) being polynomials of higher degrees are not only more difficult, but also less practical. When a strong nonlinearity is involved, it is usually desirable not to expand the characteristic for using then, say, the three first terms of the power series, when the omitted terms not really small, but to try to directly use in some way the precise elements' characteristic, such as *exp*(.), *th*(.), *arctg*(.), etc … One also notes that the nonlinearity of even a *two-term quasi-linear* characteristic is, in fact, *strong* in our problem, in the sense that $\eta$ was always found to be much smaller than the "degree of the nonlinearity" of a circuit ([1] for examples). Furthermore, since the present theory does encourage one to use two-term expansions of such "power-factorized" functions as, e.g., $(.)^{\alpha}arctg(.)$, the nonlinearity may be made very strong, from any point of view, without many-degree expansions. We thus shall limit ourselves here to some qualitative arguments, which justify, nevertheless, including such more general polynomials in the topic of analytical superposition. Remarkably, the degree of the polynomial is just the matter of how many $\alpha_p$-realizations of the digraph have to be *f*-connected. One can think here about some partial *f*-connections (only some of the respective nodes to be short-circuited), making them in a different order as regards the involvement of the different circuits. This gives some interesting "design freedom" to the theoretical consideration of the *a-test* for a high-polynomial *f*(.), especially for computer simulations. (One can even think about random connections.) We can proceed in this direction very little.

### 8.1. *The case of P = 3*

Consider now *f*-connection with $\{\alpha_1, \alpha_2, \alpha_3\}$, such that

$$\alpha_1 < \alpha_2 < \alpha_3 .$$

For the separated $\alpha$-circuits, the respective nodal potentials are $v_k(\alpha_1)$, $v_k(\alpha_2)$ and $v_k(\alpha_3)$. Because of the monotonicity of the functions $v_k(\alpha)$ in any $\alpha$-circuit, we have, for any *k*, that either

$$v_k(\alpha_1) < v_k(\alpha_2) < v_k(\alpha_3) , \qquad (30)$$

or, to the contrary,

$$v_k(\alpha_1) > v_k(\alpha_2) > v_k(\alpha_3) .$$

Both cases are treated similarly; we take for certainty (30) for the branch voltages near **b**

$$v_{k_{s''}}(\alpha_1) < v_{k_{s''}}(\alpha_2) < v_{k_{s''}}(\alpha_3) \qquad (31)$$

which results in:

$$v_{s''}(\alpha_1) < v_{s''}(\alpha_2) < v_{s''}(\alpha_3) . \qquad (32)$$

We can obtain the *f*-connection of the three given $\alpha$-circuits in two steps. We first connect only those with $\alpha_1$ and $\alpha_3$. According to (7), we obtain a circuit with $\{v_{s''}\}$ satisfying the conditions

$$v_{s''}(\alpha_1) < v_{s''} < v_{s''}(\alpha_3) , \qquad (33)$$

and an analogous double inequality takes place also for $\{v_{k_{s''}}\}$.

Comparing (33) with (32), we see that $v_{s''}$ should be relatively close to $v_{s''}(\alpha_2)$ (and, more generally, all of the $v_k$ should be relatively close to $v_k(\alpha_2)$). If so, when completing the *f*-connection of the $\alpha_1$ and $\alpha_3$ circuits by the $\alpha_2$-circuit as well, the currents will not be changed significantly. We conclude that for the circuit with $P = 3$ the analytical superposition may be of a good precision, though here quantitative analysis of the closeness of the respective potentials is much more problematic than for the *f*-connection of only two $\alpha$-circuits.

### *8.2 The case of P = 4*

Consider now the connection with $\{\alpha_1, \alpha_2, \alpha_3, \alpha_4\}$ such that

$$\alpha_1 < \alpha_2 < \alpha_3 < \alpha_4 .$$

By the same arguments as in the previous case of $P = 3$, we come now, instead of (30), to the inequalities

$$v_k(\alpha_1) < v_k(\alpha_2) < v_k(\alpha_3) < v_k(\alpha_4). \qquad (34)$$

In this case, we can perform the "*f*-connection" by first creating two *f*-circuits as follows: one is composed of the given $\alpha_1$ and $\alpha_4$ circuits, and the other of the $\alpha_2$ and $\alpha_3$ circuits. According to (23-24), we obtain, in each case, intermediate values of the respective nodal potentials. This may result in relatively very close values of the respective nodal potentials in the two *f*-circuits thus obtained, and the "*f*-connecting" of these circuits to the complete *f*-connection of the four initially given circuits will not change the currents significantly, leading here too to small errors in the analytical superposition.

We thus can essentially use the problem considered in detail above, with $P = 2$ in order to analyze the problems with $P = 3$ and $P = 4$.

### *8.3 Use of the continuity of the dependence of $v_k$ on the parameters of the type D*

Considering different special cases, one notes that a special term $D(.)^\alpha$ in an $f(.)$ may be providing the high precision of the analytical superposition (or of the $\alpha$-test). It is easy to prove (actually, it is obvious) that for the characteristic $f(.) = (.)^{\alpha_1} + D(.)^{\alpha_2}$ we can, by making $D \gg 1$, or $D \ll 1$, obtain, respectively, $v_k$ of the *f*-circuit close to that of the $\alpha_1$-circuit, or the $\alpha_2$-circuit, respectively. This may be helpful in adjusting the respective $v_k$ (or $v_{s''}$) in one of two *f*-circuits appearing at a stage of the complete *f*-connection, to the $v_k$ of the other circuit. Then, "*f*-connecting" these two circuits may be associated with very precise analytical superposition, for a wide range of $v_{in}$.

It is clear from the procedures of Subsections 2.1 and 2.2 that use of *one* such adjustable parameter is sufficient for both the $P = 3$, and $P = 4$ cases. An interesting possibility would be to *specially add*, for the case of $P = 3$, the (fourth) *small* term $D(.)^{\alpha_4}$ (with a small $D$) and a proper $\alpha_4$, trying to improve the superposition in terms of $\alpha_1$, $\alpha_2$ and $\alpha_3$.

In other words, there may be *certain* circuits with high-degree $f(.)$ for which the analytical superposition is very precise.

However, the cases of $P = 3$ and $4$ require separate detailed investigation, that could not yet be done, and we have to limit ourselves to these brief remarks.

## 9. Conclusions and final remarks

The foundation for the analytical superposition and the "conversed" $\alpha$-test, described in [1] is presented. Outside the special cases, considered in Sections 4 and 5 of [1], when the precision of the analytical superposition automatically comes out to be either ideal or asymptotically ideal, the precision of the superposition is explained by some circuit self-regulation by means of the nodal potentials that are properly changed with the *f*-connection that creating the polynomial circuit by *f*-connecting the proper $\alpha$-circuits.

The present work also presents the *f*-connection to be per se an interesting object for study, because of the possibility of applying Tellegen's theorem and because of the possibility to come to the *f*-connection by different steps, which may be useful in analytical consideration.

The author hopes that the empirical and theoretical study described in both parts of this work will be found interesting by circuit theoreticians and design specialists dealing with grid-type circuits. As regards further development of the theory, it would be most interesting, I think, if one could find applications for $\alpha$-circuits with non-real $\alpha$.

**Appendix**: **The nodes *s′′***

In the description of the *f*-circuits, index $k$ labels the nodes, and index $s$ the branches. We can write KCL at the input node, expressing the input current $i_{in} = F(v_{in})$ via the internal currents that are close to the input. See Fig. 5. For this we label the branches that enter node **a** by *s'*, and those entering node **b** by *s''*. In agreement with these notations, the *node* that is directly connected to **b** by a certain branch *s''* will be denoted by subscript $k_{s''}$. If there is a conductor directly connecting **a** and **b**, then **a** belongs to the nodes $\{k_{s''}\}$. Such a conductor is presented in $F(v_{in})$ by the term $f(v_{in})$.

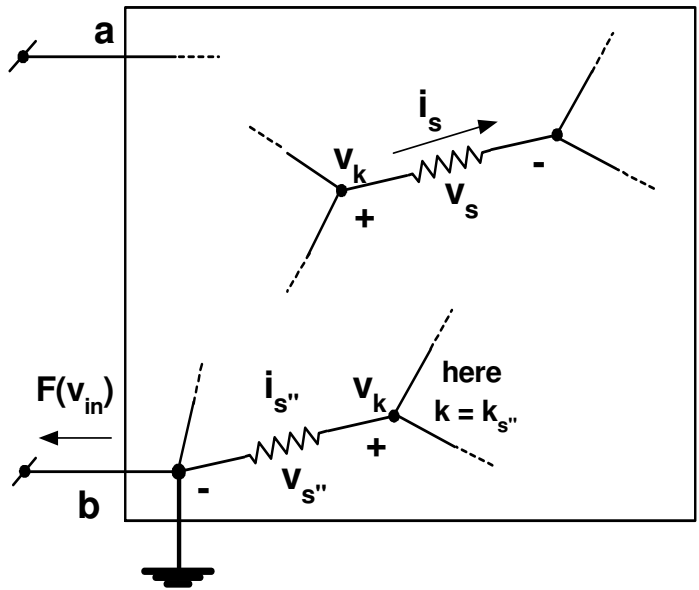

Fig 5. The branches and nodes of the *f*-circuit. Near node **b**, branches indices are not *s*, but *s''*. Observe that for the nodes close to **b**, $v_{k_{s''}} = v_{s''}$.

An expression for $F(v_{in})$ obtained by means of an input KCL equation, can employ the potentials of either only the nodes $k_{s'}$, or only $k_{s''}$. Since **b** is grounded, it is most appropriate to use the internal currents combined at **b**, and we shall prefer $k_{s''}$. Thus, the nodal voltages $v_{k_{s''}} = 0 + v_{s''} = v_{s''}$ are employed throughout the work.

Thus, the input KCL equation, written at **b** is:

$$F(v_{in}) = \sum_{s''} f(v_{s''}),$$

where the dependence on $v_{in}$ comes via that of $v_{s''}$. Usually, this equation includes, very few terms, sometimes even one term.